\documentclass[prl,twocolumn]{revtex4}
\usepackage{dcolumn}
\usepackage{multirow}
\usepackage{graphicx}
\usepackage{amssymb}
\usepackage{bm}
\usepackage{hyperref}
\usepackage{epstopdf}
\usepackage{color}
\usepackage{mathrsfs}
\usepackage{amsmath,amssymb,amsthm}
\usepackage{rotating}
\usepackage{sverb, longtable}
\usepackage{subfigure}

\usepackage[graphicx]{realboxes}
\usepackage{adjustbox}
\begin{document}

\title{ACT DR6 Leads to Stronger Evidence for Dynamical Dark Matter}
\author{Deng Wang}
\email{dengwang@ific.uv.es}
\affiliation{Instituto de F\'{i}sica Corpuscular (CSIC-Universitat de Val\`{e}ncia), E-46980 Paterna, Spain}
\author{Kazuharu Bamba}
\email{bamba@sss.fukushima-u.ac.jp}
\affiliation{Faculty of Symbiotic Systems Science, Fukushima University, Fukushima 960-1296, Japan}

\begin{abstract}
Dark matter is fundamental to the composition, structure, and evolution of the universe. Combining the ACT's cosmic microwave background, DESI's baryon acoustic oscillations with DESY5 type Ia supernova observations, we find a $3.4\,\sigma$ evidence for dynamical dark matter (DDM) with an equation of state, $\omega_{dm}(a)=\omega_{dm0}+\omega_{dma}(1-a)$. Independent of the Planck measurements, the ACT data confirms the linear relation $\omega_{dma}=-\omega_{dm0}$, inducing that the equation of state of dark matter is directly proportional to the scale factor $a$. Furthermore, the effects of DDM on the large-scale structure observables are thoroughly studied. Our findings are of great significance for understanding cosmic acceleration, structure growth, and the fate of the universe.


\end{abstract}
\maketitle

{\it Introduction.} The cosmic microwave background (CMB) radiation supporting the Big Bang theory of the universe's origin was first discovered accidentally in 1965 \cite{Penzias:1965wn} and interpreted as a remnant of the early universe in Ref.~\cite{Dicke:1965zz}. In 1992, the cosmic background explorer (COBE) satellite provided the first detailed maps of the CMB \cite{COBE:1992syq}, confirming the Big Bang theory and detecting the anisotropies in the CMB at the level of $10^{-5}$, which are believed to be the seeds of the large-scale structure (LSS) we observe in the universe today, such as galaxies and clusters of galaxies. In the late 1990s, observations of distant Type Ia supernovae (SN) revealed that the expansion of the universe is accelerating, leading to the discovery of dark energy (DE) \cite{SupernovaSearchTeam:1998fmf,SupernovaCosmologyProject:1998vns}. These observations have helped us initially establish the standard cosmology, i.e., the so-called $\Lambda$-cold dark matter ($\Lambda$CDM) model \cite{Hu:2001bc,Weinberg:2013agg}. In the first two decades of the twenty-first century, the Wilkinson Microwave Anisotropy Probe (WMAP) \cite{WMAP,WMAP:2012fli} greatly improved the resolution of CMB maps, providing detailed information about the age, composition, and shape of the universe. The Planck satellite \cite{Planck,Planck:2018vyg} further refined these measurements, offering even higher resolution and sensitivity, and confirming the $\Lambda$CDM model with an unprecedented precision. Since the first measurements of baryon acoustic oscillations (BAO) as a standard ruler by the Sloan Digital Sky Survey (SDSS) \cite{SDSS:2005xqv} and the Two-Degree Field Galaxy Redshift Survey (2dFGRS) in 2005 \cite{2dFGRS:2005yhx}, the accumulated BAO observations have independently demonstrated the validity of the $\Lambda$CDM model \cite{SDSS:2005xqv,2dFGRS:2005yhx,Beutler:2011hx,eBOSS:2020yzd,deCarvalho:2017xye,eBOSS:2017cqx,DESI:2024mwx,DESI:2024uvr,DESI:2024lzq}. However, this model is imperfect. Besides confronting two intractable challenges, namely the cosmological constant conundrum \cite{Weinberg:1988cp,Carroll:2000fy,Peebles:2002gy,Padmanabhan:2002ji} and the coincidence problem \cite{Steinhardt:1997,Zlatev:1998tr}, it also suffers from the emergent cosmic tensions, such as the Hubble constant ($H_0$) tension, matter fluctuation amplitude ($S_8$) discrepancy, and the lensing anomaly, and small-scale crisis \cite{DiValentino:2020vhf,DiValentino:2020zio,DiValentino:2020vvd,Abdalla:2022yfr,DiValentino:2025sru}, e.g., the core-cusp problem \cite{Flores:1994gz,Moore:1994yx}, diversity problem \cite{Oman:2015xda}, ``too-big-to-fail'' problem \cite{Boylan-Kolchin:2011qkt} and the satellite plane problem \cite{Kroupa:2004pt}. In recent years, the field of cosmology has seen a significant focus on DE, which has led to a relative oversight of the phenomenological possibilities of dark matter (DM) on cosmic scales. 

Historically, the notion of DM arose from inconsistencies between the dynamics of celestial systems and the expectations of Newtonian mechanics \cite{Zwicky:1933gu,Rubin:1970zza}, considering only observable matter. Initial clues about DM were derived from the examination of galaxy clusters and rotational velocities within galaxies, suggesting that luminous matter was insufficient to explain the gravitational phenomena observed. Over time, the theoretical understanding of DM has advanced. In the 1980s, the CDM paradigm was introduced, proposing that DM consists of particles that are ``cold'', or slow-moving relative to light speed. It successfully explains large-scale cosmic structures like galaxies and galaxy clusters \cite{Blumenthal:1984bp}. Other hypotheses have been put forward, including warm dark matter (WDM), which posits particles that travel at speeds approaching that of light but are still substantial enough to aid in the creation of cosmic structures. Conversely, hot dark matter (HDM) involves particles that move almost as fast as light and are less capable of facilitating the formation of such structures \cite{Dodelson:1993je}.

In astrophysics and cosmology, evidence for dark matter (DM) is derived from multiple sources, such as gravitational lensing \cite{Heymans:2012gg,Hildebrandt:2016iqg,Planck:2018lbu,DES:2017qwj}, galaxy clustering \cite{DES:2017myr,DES:2021wwk,DES:2025xii}, CMB \cite{COBE:1992syq,WMAP:2012fli,Planck:2018vyg}, and the behavior of galaxy clusters and individual galaxies \cite{Weinberg:2013agg}. Especially, gravitational lensing, where light bends around massive objects, directly reveals the DM distribution in the universe. The precise Planck CMB data \cite{Planck,Planck:2018vyg} have disclosed that DM constitutes approximately 85\% of the total matter in the universe, playing a crucial role in shaping its evolution and dynamics. In particle physics, the quest for dark matter (DM) has spurred various detection initiatives, encompassing direct and indirect detection experiments, as well as collider searches. Direct detection efforts focus on  identifying the rare interactions between DM particles and ordinary matter. Experiments like XENON \cite{XENON,XENON100:2012itz}, LUX \cite{LUX,LUX:2016ggv}, and PandaX \cite{PANDAX,PandaX:2014mem} utilize substantial quantities of liquid xenon or other substances to detect faint signals from these interactions \cite{XENON:2018voc}. Indirect detection methods seek evidence of DM through its decay or annihilation byproducts, such as gamma rays, neutrinos and antimatter. Instruments like the space-based Fermi-LAT telescope \cite{FERMI-LAT} and ground-based observatories such as HESS \cite{HESS} and MAGIC \cite{MAGIC,MAGIC:2016xys,Profumo:2017obk} are employed to identify these signals from cosmic sources \cite{Klasen:2015uma,Gaskins:2016cha}. Collider searches, notably at the Large Hadron Collider (LHC) \cite{LHC}, aim to generate DM particles and identify their presence through missing energy and momentum signatures. These experiments complement direct and indirect detection strategies, potentially uncovering new physics beyond the Standard Model. 

Up to now, DM has yet to be detected in astronomical observations and particle experiments, presenting substantial challenges for both theoretical and experimental physics. In light of the fact that the potential roles and characteristics of DM at a phenomenological level have not received as much attention as DE, in this study, we will focus on exploring the phenomenological aspects of DM on cosmic scales. Interestingly, the DESI collaboration recently reported the substantial evidence of dynamical dark energy (DDE) \cite{DESI:2025fii}, based on their measurements of BAO in galaxy, quasar and Lyman-$\alpha$ forest tracers from the second data release (DR2) of the Dark Energy Spectroscopic Instrument (DESI) \cite{DESI:2025zgx,DESI:2025zpo}. However, this conclusion is obtained by assuming that DM is absolutely cold. This prompts us to ponder whether DM like DE is dynamical over time. A more intriguing question is whether the dark sector of the universe consists of DDM and DDE instead of CDM and the cosmological constant?

Based on the theoretical motivation proposed in Ref.~\cite{Hu:1998kj}, although the CPL-like non-constant DM EoS has been investigated in Refs.~\cite{Kumar:2012gr,Kumar:2019gfl}, the incomplete parameter space of DM EoS ($\omega_{dm0}>0$ and $\omega_{dma}>0$) significantly restricts the capture of all the possibly phenomenological effects and the exploration of new physics on cosmic scales. Interestingly, extending the theoretical parameter space of DM EoS to the full space, the robust $\sim2\,\sigma$ evidences of DDM are found using recent observations in Ref.~\cite{Wang:2025zri}. Planck CMB data support a very strong linear relation $\omega_{dma}=-\omega_{dm0}$, inducing a directly proportional DM EoS, $\omega_{dm}(a)=\omega_{dm}a$, where the $\sim2\,\sigma$ DDM evidences are well captured and even strengthened. We demonstrate that there are beyond $2\,\sigma$ evidences of the coexistence of DDM and DDE using the joint constraint from Planck CMB, DESI DR2 BAO, and Pantheon+ SN data. In this study, using the latest ACT DR6 CMB observations, we confirm the linear relation $\omega_{dma}=-\omega_{dm0}$ independent of Planck. Interestingly, the evidence for DDM has reached a $3.4\,\sigma$ significance level revealed by the combination of ACT CMB, DESI DR2 BAO, and DESY5 SN data.

{\it Basics.} Considering a homogeneous and isotropic universe within the framework of general relativity \cite{Einstein:1916vd}, the Friedmann equations \cite{Friedman:1922kd} are written as $H^2=(8\pi G\rho)/3$ and $\ddot{a}/a=-4\pi G(\rho+3p)/3$, where $\rho$ and $p$ denote the mean energy densities and pressures of cosmic species including the radiation, baryons, DM and DE, and $H$ is the Hubble parameter at a scale factor $a$. Assuming the equation of state (EoS) of DM $\omega_{dm}(a)=\omega_{dm0}+\omega_{dma}(1-a)$ and the Chevallier-Polarski-Linder (CPL) DE EoS $\omega(a)=\omega_0+\omega_a(1-a)$ \cite{Chevallier:2000qy,Linder:2002et} and combining the above equations, the normalized Hubble parameter $E(a)\equiv H(a)/H_0$ reads as
\begin{equation}
E(a)=\sqrt{\Omega_{r}a^{-4}+\Omega_{b}a^{-3}+\Omega_{\rm DM}(a)+\Omega_{\rm DE}(a)}, \label{eq:ez}
\end{equation}
where $\Omega_{\rm DM}(a)=\Omega_{dm}a^{-3(1+\omega_{dm0}+\omega_{dma})}\mathrm{e}^{3\omega_{dma}(a-1)}$ and $\Omega_{\rm DE}(a)=\Omega_{de}a^{-3(1+\omega_0+\omega_a)}\mathrm{e}^{3\omega_a(a-1)}$, where $\Omega_r$, $\Omega_b$, $\Omega_{dm}$ and $\Omega_{de}$ ($=1-\Omega_{dm}-\Omega_{b}-\Omega_{r}$) denote the present-day radiation, baryon, DM and DE fractions, respectively. It reduces to $\Lambda$CDM when $\omega_0=-1$ and $\omega_{dm0}=\omega_{dma}=\omega_a=0$ .

{\it Data and methodology.} For reducing the computing cost, we use the ACT DR6 \texttt{ACT-lite} likelihood including the temperature-temperature (TT), temperature-E mode polarization and polarization-polarization (EE) power spectra, which have white noise levels that improve over those of Planck by roughly a factor of three in polarization and a factor of two in temperature, in the multipole range $600\leqslant\ell\leqslant8500$ and the CMB signal extracted in $600\leqslant\ell\leqslant6500$ \cite{ACT:2025fju}. 
As a comparison, we employ the Planck 2018 high-$\ell$ \texttt{plik} likelihood including the TT data in $30\leqslant\ell\leqslant2508$, EE and TE spectra in $30\leqslant\ell\leqslant1996$, and the low-$\ell$ TT \texttt{Commander} and \texttt{SimAll} EE likelihoods in $2\leqslant\ell\leqslant29$ \cite{Planck:2019nip}. We take the ACT DR6 CMB lensing in $40\leqslant\ell \leqslant763$ \cite{ACT:2023dou} and the Planck PR4 CMB lensing in $8\leqslant\ell \leqslant400$ \cite{Planck:2018lbu}. We adopt 13 DESI DR2 BAO measurements including the BGS, LRG1, LRG2, LRG3+ELG1, ELG2, QSO, and Ly$\alpha$ samples at the effective redshifts $z_{\rm eff}=0.295$, 0.51, 0.706, 0.934, 1.321, 1.484 and $2.33$, respectively \cite{DESI:2025zgx,DESI:2025fii,DESI:2025zpo}. We employ the DESY5 SN sample including 1735 effective data points in the range $0.025\leqslant z\leqslant1.130$ \cite{DES:2024jxu}. Hereafter, ``ACT'' denotes ACT DR6 CMB plus ACT DR6 CMB lensing, while ``ACT+Planck'' represents ACT DR6 CMB plus Planck CMB plus ACT DR6 CMB lensing plus Planck PR4 CMB lensing. We refer to ACT+DESI, ACT+DESI+DESY5 and ACT+Planck+DESI+DESY5 as ``AD'', ``ADS'', and ``APDS'', respectively. Same as the ACT collaboration \cite{ACT:2025tim}, in all the ACT-only analyses, we incorporate the \texttt{Sroll2} likelihood \cite{Pagano:2019tci}, which is a Planck measurement of the optical depth to reionization from the low-$\ell$ EE spectra in $\ell\leqslant30$.  

\begin{figure}[h]
	\centering
	\includegraphics[scale=0.52]{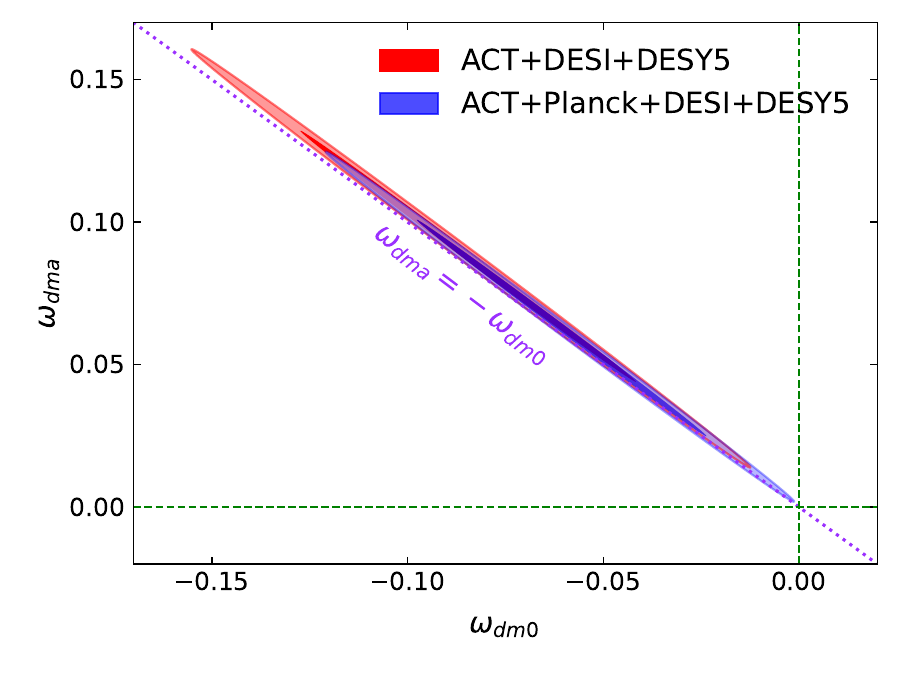}
	\caption{Two-dimensional marginalized posterior distributions of the DM EoS parameters from the combinations of ACT+DESI+DESY5 and ACT+Planck+DESI+DESY5 in the DDM model. The purple dotted line depicts the linear relation $\omega_{dma}=-\omega_{dm0}$, while the cross point of the green dashed lines denote the $\Lambda$CDM model.}\label{f1}
\end{figure}

\begin{figure}
	\centering
	\hspace*{-0.5cm}
	\includegraphics[scale=0.45]{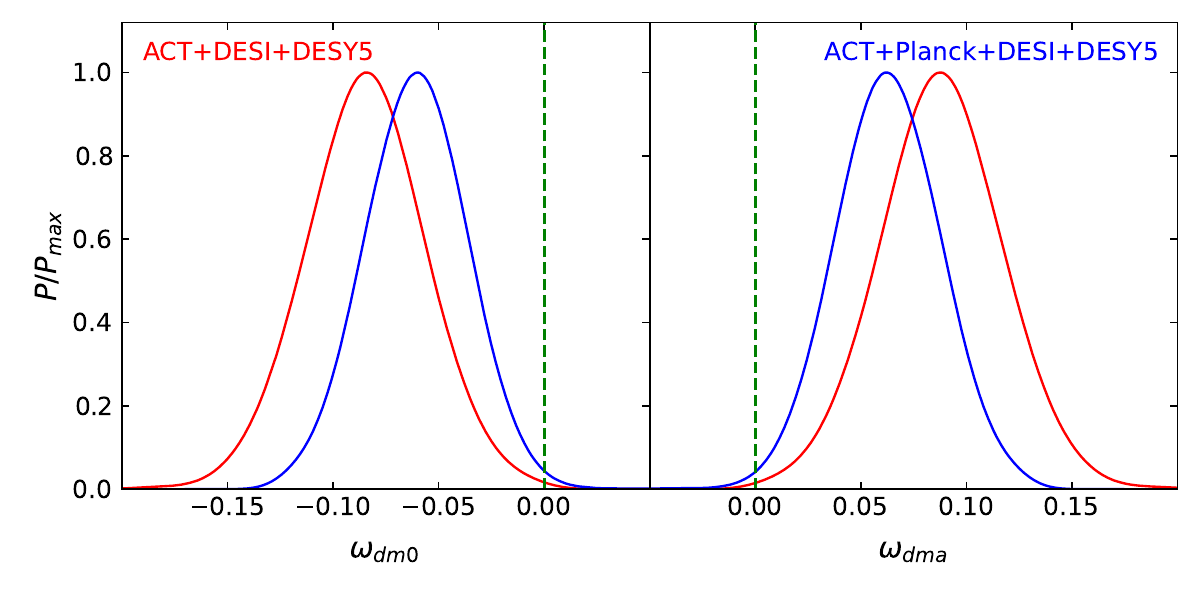}
	\caption{One-dimensional marginalized posterior distributions of the DM EoS parameters from the combinations of ACT+DESI+DESY5 and ACT+Planck+DESI+DESY5 in the DDM model. The green dashed lines denote $\omega_{dm0}=\omega_{dma}=0$ predicted by the $\Lambda$CDM model.}\label{f2}
\end{figure}

\begin{figure}
	\centering
	\includegraphics[scale=0.42]{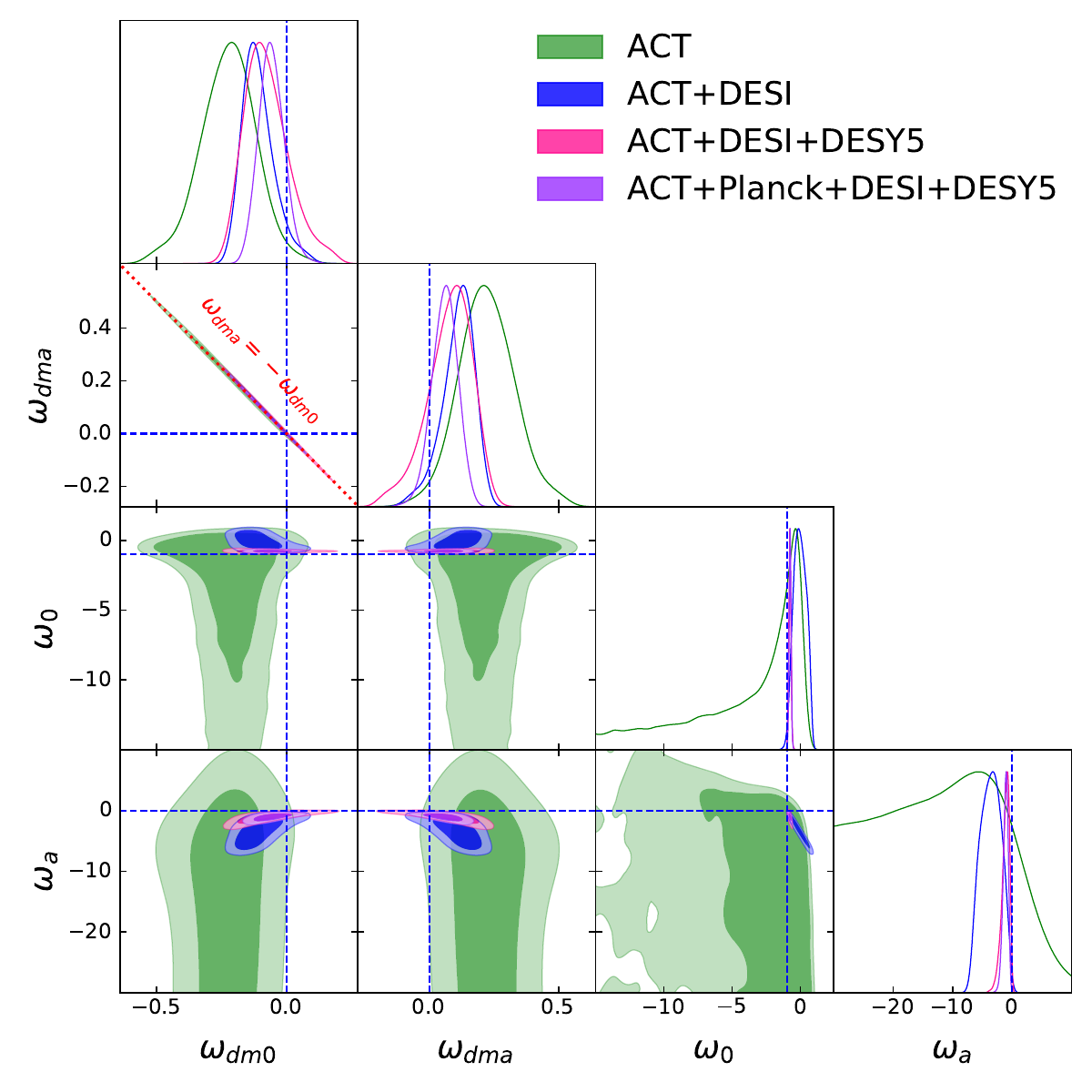}
	\caption{One-dimensional and two-dimensional marginalized posterior distributions of the DDM and DDE EoS parameters from various datasets in the DDME model. The blue dashed lines denote $\omega_{dm0}=\omega_{dma}=\omega_a=0$ and $\omega_0=-1$, while the red dotted line represents the linear relation $\omega_{dma}=-\omega_{dm0}$.}\label{f3}
\end{figure}

\begin{table*}[t!]
	\renewcommand\arraystretch{1.5}
	\caption{Mean values and $1\,\sigma$ (68\%) errors of free parameters from various datasets in the DDM, SDDM and DDME models. We quote $2\,\sigma$ (95\%) upper limits for parameters with weak constraints. }
	\setlength{\tabcolsep}{4mm}{
		\begin{tabular} { c |c| c |c| c|c }
			\hline
			\hline
			
			\multicolumn{2}{c|}{Data}                & ACT      & ACT+DESI          & ACT+DESI+DESY5    &ACT+Planck+DESI+DESY5         \\
			\hline
			\multirow{2}{1cm}{DDM}  & $\omega_{dm0}$  &$-0.36^{+0.17}_{-0.30}$ & $-0.028\pm 0.040     $ & $-0.084\pm 0.029    $ & $-0.061\pm 0.025     $                               \\
			\cline{2-6}
			&    $\omega_{dma}$  & $0.36^{+0.30}_{-0.18}    $ & $0.030\pm 0.041      $ & $0.088\pm 0.030       $ & $0.063\pm 0.025       $                             \\

			\hline
			\multirow{1}{1cm}{SDDM}  &     $\omega_{dm}$  & $-0.13\pm 0.20   $ & $0.037\pm 0.017      $ & $0.006\pm 0.015      $ & $0.007\pm 0.011          $ \\
			
			\hline
			\multirow{4}{1cm}{DDME}  &     $\omega_{dm0}$  & $-0.22\pm 0.11             $ & $-0.108^{+0.044}_{-0.073}  $ & $-0.075^{+0.065}_{-0.100}   $ & $-0.058^{+0.046}_{-0.051}  $                          \\
			\cline{2-6}
			&     $\omega_{dma}$  & $0.22\pm 0.11              $ & $0.110^{+0.074}_{-0.046}   $ & $0.078^{+0.100}_{-0.067}    $ & $0.059^{+0.052}_{-0.047}   $       \\
			\cline{2-6}                        
			&     $\omega_{0}$   & $< 0.517         $ & $-0.03\pm 0.45             $ & $-0.777^{+0.070}_{-0.082}  $ & $-0.758\pm 0.072           $                 \\	
			\cline{2-6}	                        		                        
			&     $\omega_{a}$    & $< 2.98            $ & $-3.60\pm 1.70               $ & $-1.10^{+0.80}_{-0.40}     $ & $-1.10^{+0.50}_{-0.33}     $            \\
			
			\hline
			\hline
	\end{tabular}}
	\label{t1}
\end{table*}

To compute the background dynamics and power spectra of the universe, we develop the \texttt{DDMCAMB}, a modified version of the Boltzmann code \texttt{CAMB} \cite{Lewis:1999bs}, which works with any DM EoS model and permits us to extract the possible new physics of DM on cosmic scales using the phenomenological EoS approach.  
To perform the Bayesian analysis, we take the Monte Carlo Markov Chain (MCMC) approach to infer the posterior distributions of free parameters using \texttt{Cobaya} \cite{Torrado:2020dgo}. We assess the convergence of MCMC chains using the Gelman-Rubin criterion $R-1\lesssim 0.01$ \cite{Gelman:1992zz} and analyze them using \texttt{Getdist} \cite{Lewis:2019xzd}.

We take the following uniform priors for free parameters: the baryon fraction $\Omega_bh^2 \in [0.005, 0.1]$, CDM fraction $\Omega_ch^2 \in [0.001, 0.99]$, angular acoustic scale at the recombination epoch $100\theta_{\rm MC} \in [0.5, 10]$, scalar spectral index $n_s \in [0.8, 1.2]$, amplitude of the primordial power spectrum $\ln(10^{10}A_s) \in [2, 4]$, optical depth $\tau \in [0.01, 0.8]$, today's DM EoS $\omega_{dm0} \in [-2, 2]$, amplitude of DM evolution $\omega_{dma} \in [-2, 2]$ and $\omega_{dm} \in [-2, 2]$, present-day DE EoS $\omega_0 \in [-15, 20]$ and amplitude of DE evolution $\omega_a \in [-30, 10]$. 
Hereafter, we denote the CPL-like DDM, single-parameter proportional DDM and DDM plus DDE models as ``DDM'', ``SDDM'', and ``DDME'', respectively.

{\it DDM evidence induced by ACT.} Using ACT alone, we find $\omega_{dm0}=-0.36^{+0.17}_{-0.30}$ and $\omega_{dma}=0.36^{+0.30}_{-0.18}$ implying a $1.69\,\sigma$ hint of DDM, which not only is consistent with the $\sim2\,\sigma$ evidence from Planck CMB alone \cite{Wang:2025zri} but also confirms in an independent way the directly proportional DM EoS $\omega_{dm}(a)=\omega_{dm}a$ found in Ref.~\cite{Wang:2025zri}, where $\omega_{dm}$ is the sole parameter characterizing the redshift evolution of DM EoS. The extent of deviation from zero in today's DM pressure corresponds precisely to the significance level of evolving DM. Interestingly, ADS gives $\omega_{dma}=0.088\pm 0.030$ indicating a $3.40\,\sigma$ evidence of DDM (see Table~\ref{t1}), while AD just gives a small preference of $\omega_{dma}>0$. However, the addition of Planck CMB to ADS provides $\omega_{dma}=-0.061\pm 0.025$ reducing the significance to $2.98\,\sigma$. ADS and APDS further help confirm the proportional DM EoS (see Fig.~\ref{f1}) and exhibit larger deviations (see Fig.~\ref{f2}) from $\Lambda$CDM than those in Ref.~\cite{Wang:2025zri}. It is noteworthy that exponential quintessence can well explain such a redshift evolution of DM EoS on cosmic scales \cite{Wang:2025rll,Chen:2025wwn}. 
For the case of SDDM, ACT-only constraint only provides a small preference of $\omega_{dm}<0$ unlike Planck \cite{Wang:2025zri}. Interestingly, AD gives $\omega_{dm}=0.037\pm 0.017$ indicating a $2.18\,\sigma$ hint of $\omega_{dm}>0$, which, at first glance, seems inconsistent with the parameter space obtained in DDM. Actually, this anomalous positive value of $\omega_{dm}$ sources from the fact that DESI plus the high-precision information of comoving sound horizon $r_d$ from ACT leads to a larger $H_0$ than ACT. Since $\omega_{dm}$ is positively correlated with $H_0$ in SDDM, a larger $H_0$ gives a more positive $\omega_{dm}$. 
The constraints on $\omega_{dm}$ from ADS and APDS are more precise but just show a slight preference of $\omega_{dm}>0$. Intriguingly, ACT alone gives a tighter constraint on $\omega_{dma}=0.22\pm 0.11$ in DDME, implying a $2\,\sigma$ signal of DDM, than $\omega_{dma}=0.36^{+0.30}_{-0.18}$ in DDM. The origin of this unusual constraint is that DDME provides a much tighter $\Omega_ch^2$ and $\theta_{\rm MC}$ than DDM after considering the evolution of DE over time. This reveals the preference of ACT data for the dynamical nature of the dark sector of the universe. Notice that ACT gives a weaker constraint on the DE EoS parameters $(\omega_0,\,\omega_a)$ than Planck. Furthermore, we find that AD, ADS, and APDS give $\omega_{dma}=0.110^{+0.074}_{-0.046}$, $0.078^{+0.100}_{-0.067}$ and $0.059^{+0.052}_{-0.047}$ and $\omega_{a}=-3.60\pm 1.70$, $-1.10^{+0.80}_{-0.40}$ and $-1.10^{+0.50}_{-0.33}$, respectively, indicating the robust $\sim2\,\sigma$ evidences of the coexistence of DDM and DDE (see Fig.~\ref{f3}). This means that the nature of dark sector is likely DDM plus DDE instead of CDM plus the cosmological constant. More interestingly, the proportional DM EoS has once again been well confirmed in DDME.

{\it DDM versus LSS.} Since a small variation of DM EoS could significantly affect the gravitational potential, DM perturbations, and background expansion, DDM could influence the observations of cosmic LSS. As an example, here we use the LSS observations from the Dark Energy Survey Year 1 (DESY1) including three two point correlation functions, i.e., galaxy clustering, cosmic shear and galaxy-galaxy lensing, which covers the redshift range $0.2\leqslant z\leqslant1.3$ in $0.042\leqslant k\leqslant12.566$ $h\,$Mpc$^{-1}$ \cite{DES:2017qwj,DES:2017myr}. Adding DESY1 to ADS, we find $\omega_{dma}=0.039\pm 0.026$ and $\omega_{dm0}=-0.037\pm 0.025$ suggesting a $1.50\,\sigma$ hint of DDM. Even if considering the low-redshift LSS observations, the strong linear relation $\omega_{dma}=-\omega_{dm0}$ still holds in DDM. The reason why the significance of DDM reduces from $3.40\,\sigma$ to $1.50\,\sigma$ is that DESY1 prefers a lower CDM fraction in DDM than those in $\Lambda$CDM, which induces a larger $H_0$ and consequently reduces the significance level of DDM. Since adding DESY1 introduces clear inconsistencies among datasets and more complexities in data analyses, we exclude DESY1 in our data combinations (see Table~\ref{t1}). 
However, in order to probe the possible deviation from $\Lambda$CDM with future high-precision LSS data, it is important to study theoretically the effects of DDM on the typical LSS observables or quantities, because DM plays a crucial role in shaping the evolution and structure formation of the universe. We therefore consider the CMB lensing potential power spectrum (PS), Weyl potential PS, velocity PS, density-velocity cross PS, relative baryon-CDM velocity PS, correlation function between CMB lensing and galaxy number counts, besides the matter PS and the angular PS of CMB (see the supplementary material). We find that an evolving DM EoS, especially the negative pressure DM induced by the ACT DR6 observations, can significantly influence the motion of cosmic species and the evolution of LSS. The allowed DDM parameter space by the tight constraints from ADS and APDS could lead to rich LSS phenomena deviated from the standard prediction, suggesting the novel possibilities of fundamental physics.  

{\it Discussions and conclusions.} As a cross check of Plank dominated constraints \cite{Wang:2025zri}, ACT confirms the validity of the strong linear relation $\omega_{dma}=-\omega_{dm0}$. Up to now, since all the CMB based datasets support this simple relation, we propose a conjecture that DM obeys the directly proportional DM EoS $\omega_{dma}=\omega_{dm}a$ emerged on cosmic scales. However, the combination of CMB (ACT or Planck) and DESI gives biased constraints, since $r_d$ derived from CMB induces a larger $H_0$. As a special case of DDM, $\Lambda$CDM with a varying constant DM EoS $\omega_{dm0}$ have been constrained with recent cosmological observations \cite{Wang:2025zri,Muller:2004yb,Calabrese:2009zza,Xu:2013mqe,Thomas:2016iav,Kopp:2018zxp,Kumar:2025etf}. Although the uncertainties of $\omega_{dm0}$ is even limited to be $\sim \mathcal{O}(-4)$, we emphasize that this scenario cannot interpret the nature of DM on cosmic scales from the viewpoint of DM EoS. 

An alternative way to study the low-redshift evolution of DM EoS is the model-independent Gaussian processes (GP) \cite{Wang:2025zri,Abedin:2025dis}. However, unfortunately, aside from the inability to incorporate the perturbation information, inserting four quantities $H_0$, $\Omega_{m}$, $\Omega_{b}$ and $r_d$ largely reduces the model independence of the GP reconstruction and consequently affect the validity of reconstructing results. Moreover, the standard GP approach suffers from some internal problems such as stationary assumption, smoothness bias, overfitting risk, extrapolation weakness and lack of interpretability \cite{Rasmussen:2004}. 

In principle, DDM could be realized by modified gravity (e.g., $f(R)$ gravity \cite{Sotiriou:2008rp,DeFelice:2010aj,Nojiri:2010wj,Capozziello:2011et,Nojiri:2017ncd}) on cosmic scales via the matter-geometry degeneracy revealed by the Einstein's field equation \cite{Einstein:1916vd}, because both can affect the background dynamics and the perturbation evolution of the universe. However, if pursing a more consistent theory with the observations of galactic rotation curves \cite{Kent:1987zz,Milgrom:1988,Begeman:1991iy,Sanders:1996ua} and baryonic Tully-Fisher relation \cite{McGaugh:2000sr,McGaugh:2005qe}, one should introduce the Tensor-Vector-Scalar (TeVeS) theory \cite{Sanders:1996wk,Bekenstein:2004ne}, which could reproduce the key cosmological and astrophysical observables \cite{Skordis:2020eui}.  

Interestingly, DDM can not only resolve the neutrino mass tension between cosmological observations and terrestrial experiments, but also challenge the prevailing understanding of cosmic acceleration and deepen our insight into the universe's evolution \cite{Wang:2025zri}. The observational evidence of the phenomenological DDM on cosmic scales could be a smoking gun of modified gravities or novel particle physics scenarios. Future high-precision cosmological surveys such as SKA \cite{SKA,SKA:2018ckk} can further help compress the parameter space of DM EoS and reveal the nature of dark sector of the universe.      

{\it Acknowledgements.} DW is supported by the CDEIGENT fellowship of the Consejo Superior de Investigaciones Científicas (CSIC). KB is supported by the JSPS KAKENHI Grant Number 24KF0100 and Competitive Research Funds for Fukushima University Faculty (25RK011).

\clearpage

\appendix

\onecolumngrid
\section{\large Supplementary Material}
\twocolumngrid

\onecolumngrid

\setcounter{equation}{0}
\setcounter{figure}{0}
\setcounter{table}{0}

\section*{A. ACT-only constraint on DDM}
\begin{figure}[h]
	\centering
	\includegraphics[scale=0.52]{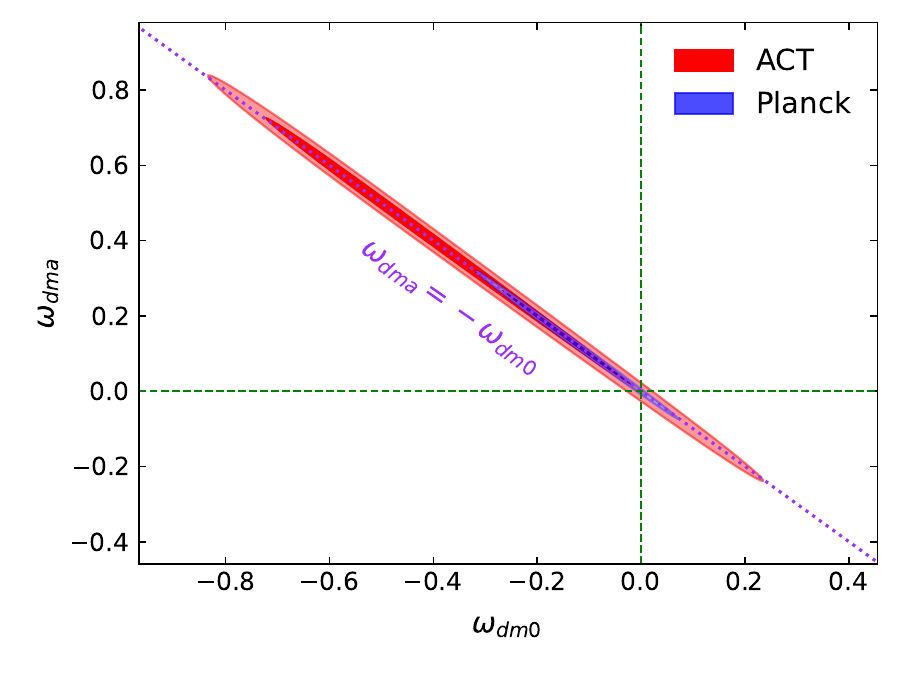}
	\caption{Two-dimensional marginalized posterior distributions of the DM EoS parameters from the ACT and Planck CMB observations in the DDM model. The purple dotted line depicts the linear relation $\omega_{dma}=-\omega_{dm0}$, while the cross point of the green dashed lines denote the $\Lambda$CDM model.}\label{fs1}
\end{figure}

The ACT-only constraint gives $\omega_{dm0}=-0.36^{+0.17}_{-0.30}$ and $\omega_{dma}=0.36^{+0.30}_{-0.18}$ indicating a $1.69\,\sigma$ hint of DDM, which not only is compatible with the $\sim2\,\sigma$ evidence from the Planck-only constraint  but also independently confirms the proportional DM EoS $\omega_{dm}(a)=\omega_{dm}a$ found in Ref.~\cite{Wang:2025zri}. In Fig.~\ref{fs1}, we show the constraining consistency between ACT and Planck CMB observations in the $\omega_{dm0}$-$\omega_{dma}$ plane.

\section*{B. Special case: constraints on $\omega_{dm0}$}
Here we denote the scenario of $\Lambda$CDM with a varying constant DM EoS $\omega_{dm0}$ as ``$\Lambda\omega$DM''. Confronting this model with observations, we find that ACT gives $\omega_{dm0}=-0.0028^{+0.0031}_{-0.0036}$, which is well consistent with $\omega_{dm0}=-0.0015\pm 0.0018$ given by Planck within the $1\,\sigma$ confidence level. Notice that they both prefer slightly a negative pressure DM. The constraints $\omega_{dm0}=0.00134\pm 0.00046$, $0.00104\pm 0.00046$, and $0.00054\pm 0.00029$ from AD, ADS, and APDS provide a $3.33\,\sigma$, $2.25\,\sigma$, and $1.86\,\sigma$ evidence of $\omega_{dm0}>0$, respectively (see Table.~\ref{ts1}). These anomalous values are not unexpected because a larger $H_0$ is induced by the combination of new DESI BAO measurements and the derived $r_d$ from CMB. It is interesting that considering an evolving DM leads to the fact the constraint on today's DM EoS $\omega_{dm0}$ in the DDM model is clearly weakened by $\sim2$ orders compared to that in the $\Lambda\omega$DM model.

\begin{figure}[h!]
	\centering
	\includegraphics[scale=0.55]{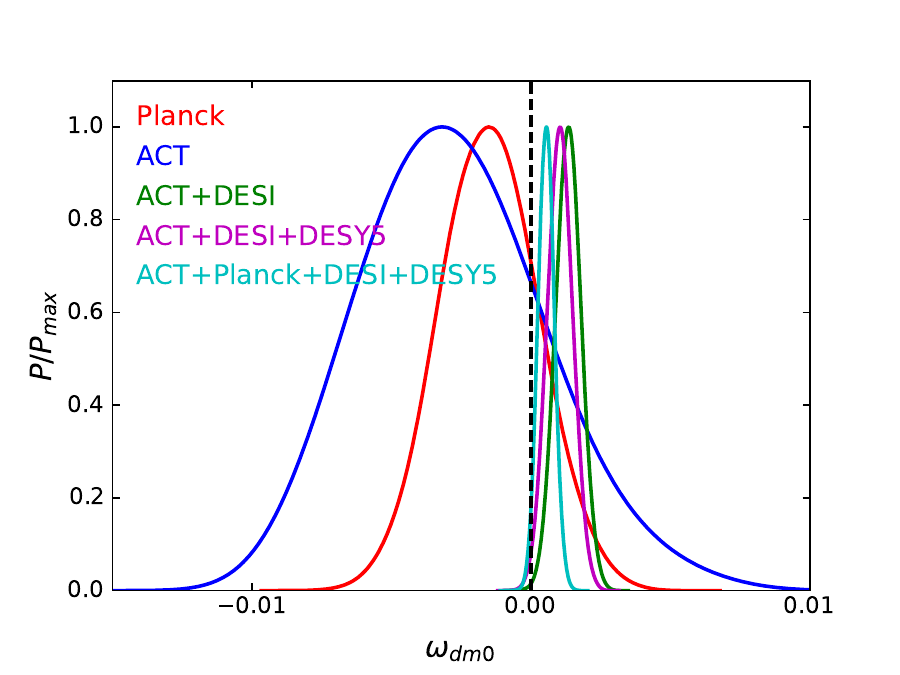}
	\caption{One-dimensional marginalized posterior distributions of the parameter $\omega_{dm0}$ from the CMB based datasets in the $\Lambda\omega$DM model. The black dashed line denotes the $\Lambda$CDM model.}\label{fs2}
\end{figure}

\begin{table*}[h!]
	\renewcommand\arraystretch{1.6}
	\begin{center}
		\caption{Mean values and $1\,\sigma$ (68\%) errors of the parameter $\omega_{dm0}$ from different datasets in the $\Lambda\omega$DM model. }
		\setlength{\tabcolsep}{10mm}{
			\begin{tabular}{l |c }
				\hline
				\hline
				Parameter & $\omega_{dm0}$    \\
				\hline 
				Planck &  $-0.0015\pm 0.0018 $                     \\
				ACT &  $-0.0028^{+0.0031}_{-0.0036} $                     \\
				ACT+DESI &  $0.00134\pm 0.00046$        \\
				ACT+DESI+DESY5 &  $0.00104\pm 0.00046$    \\
				ACT+Planck+DESI+DESY5 &  $0.00054\pm 0.00029$      \\				
				\hline
				\hline
		\end{tabular}}
		\label{ts1}
	\end{center}
\end{table*}

\section*{C. The effects of DDM on LSS}

\begin{figure}[h!]
	\centering
	\includegraphics[scale=0.58]{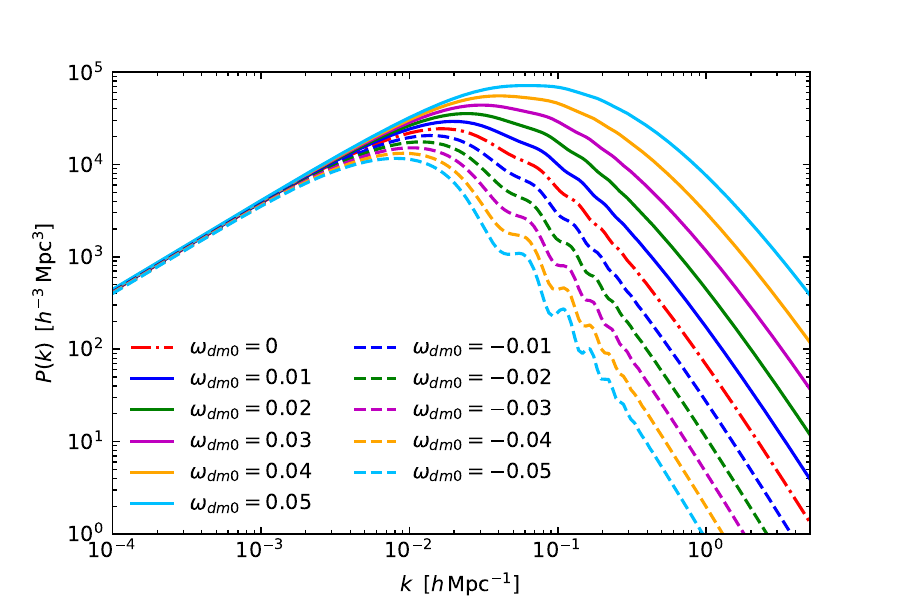}
	\includegraphics[scale=0.58]{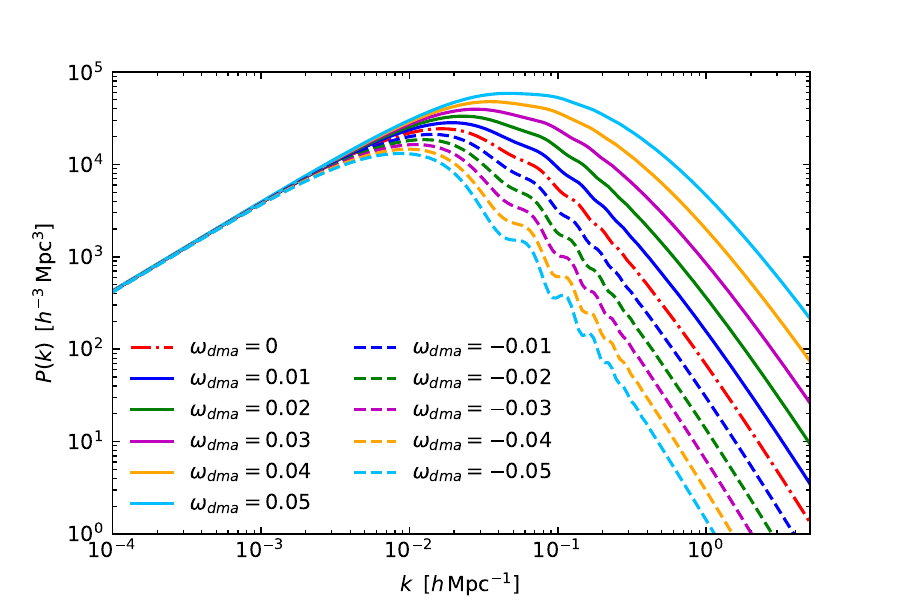}
	\includegraphics[scale=0.58]{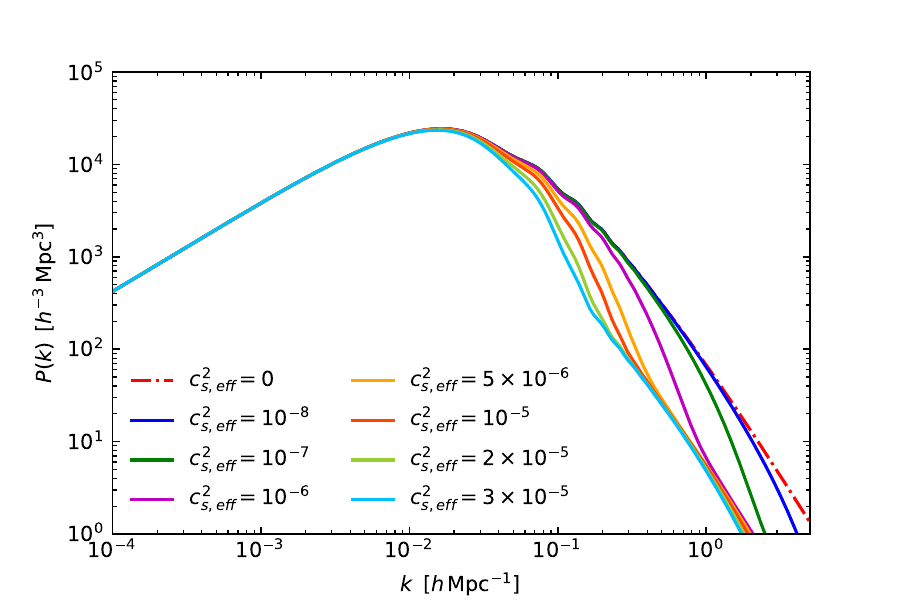}
	\includegraphics[scale=0.58]{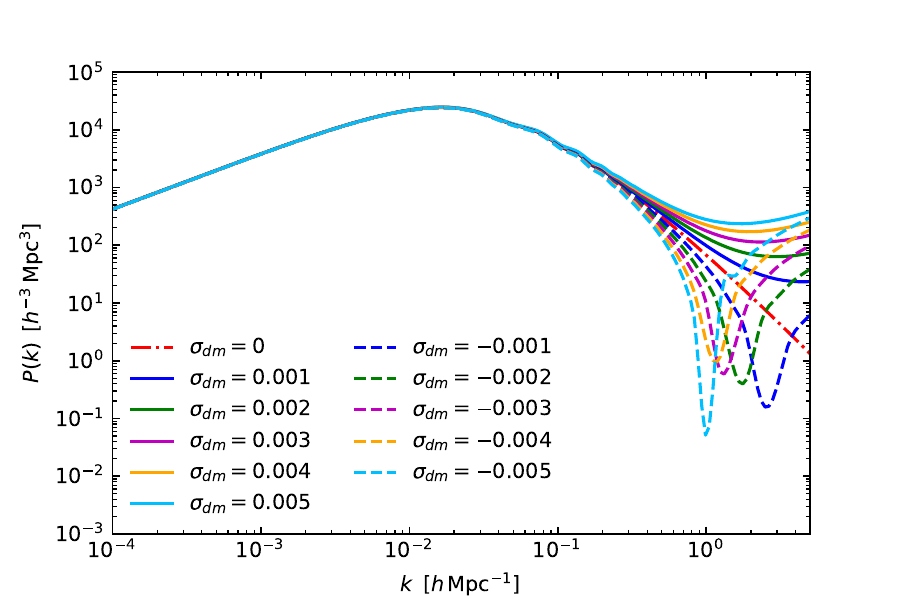}
	\caption{The matter power spectra derived from different values of $\omega_{dm0}$, $\omega_{dma}$, $c^2_{s,eff}$, and $\sigma_{dm}$  in the DDM model, respectively.}\label{fs3}
\end{figure}

\begin{figure}[h!]
	\centering
	\hspace*{-2cm}
	\includegraphics[scale=0.52]{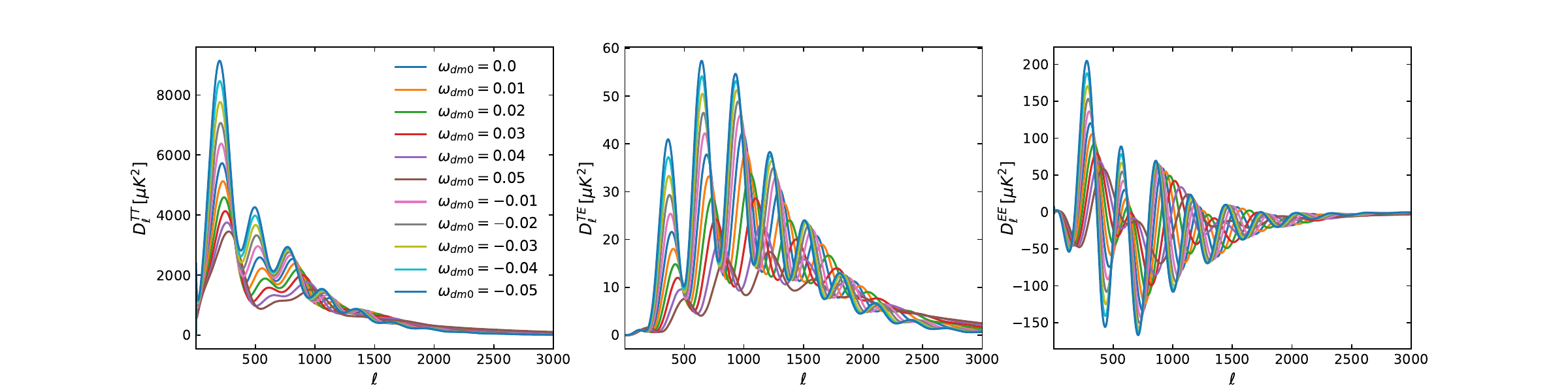}
	\hspace*{-2cm}
	\includegraphics[scale=0.52]{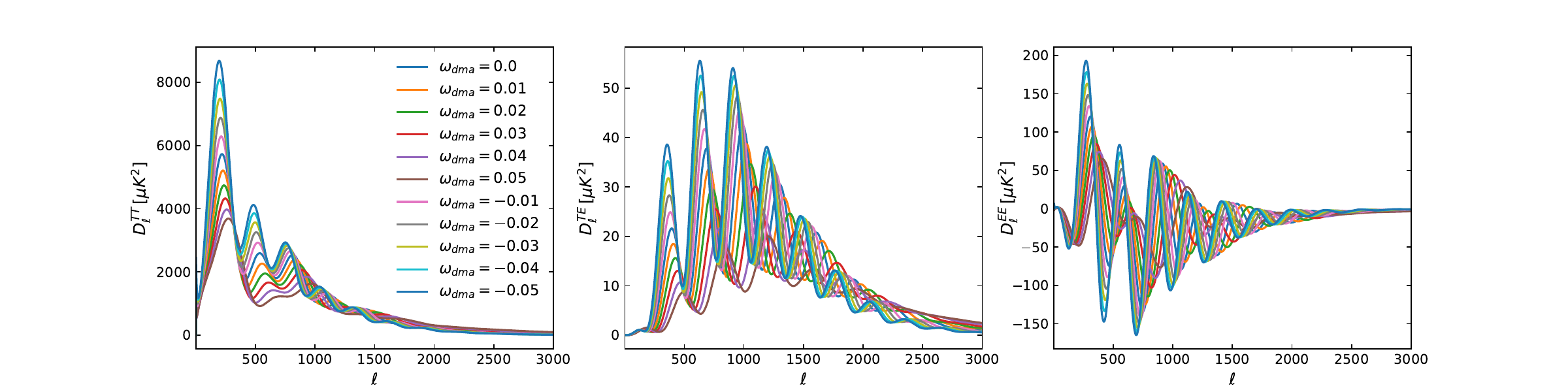}
	\hspace*{-2cm}
	\includegraphics[scale=0.52]{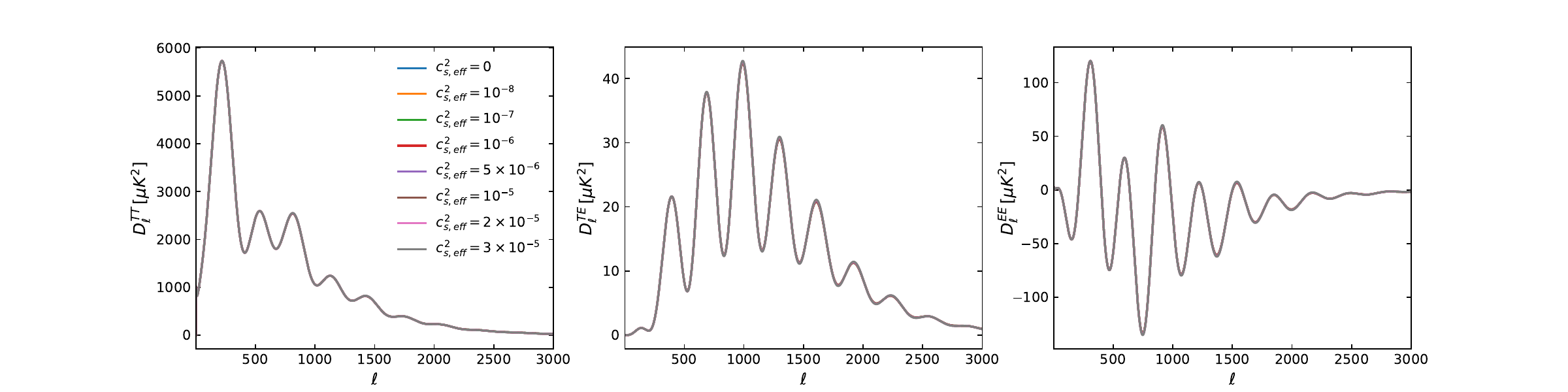}
	\hspace*{-2cm}
	\includegraphics[scale=0.52]{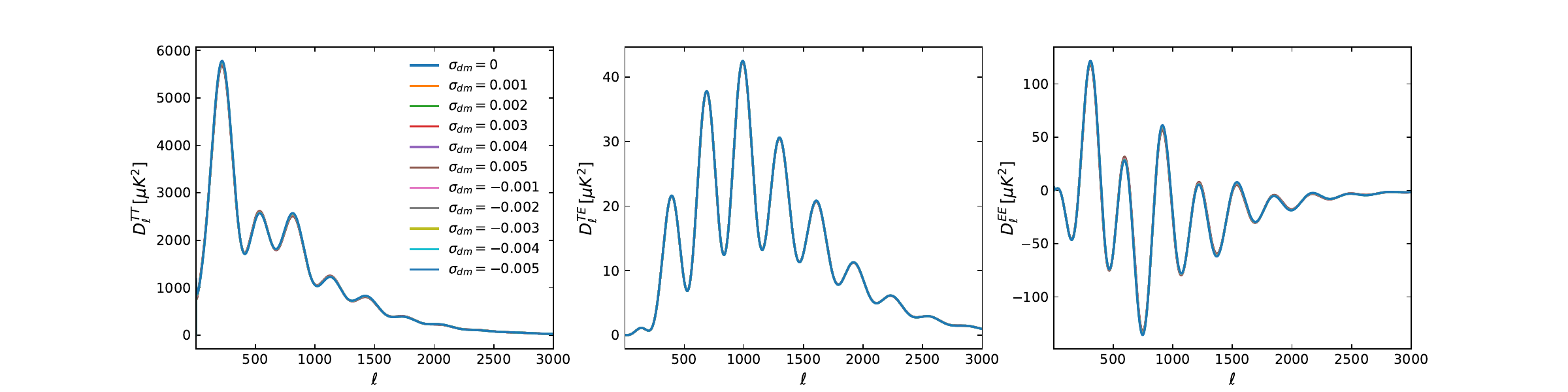}
	\caption{The CMB temperature, E-mode polarization, and temperature-E-mode polarization power spectra derived from different values of $\omega_{dm0}$, $\omega_{dma}$, $c^2_{s,eff}$, and $\sigma_{dm}$  in the DDM model, respectively.}\label{fs4}
\end{figure}

In the synchronous gauge under the framework of general relativity \cite{Mukhanov:1992,Ma:1995,Malik:2009},  the perturbation evolution of DM as a perfect fluid read as
\begin{align}
\dot{\delta}_{dm}&=-(1+\tilde{\omega}_{dm})(\theta_{dm}+\frac{\dot{h}}{2})+\frac{\dot{\tilde{\omega}}_{dm}}{1+\tilde{\omega}_{dm}}\delta_{dm}-3\mathcal{H}(c^2_{s,eff}-c^2_{s,ad})\left[\delta_{dm}+3\mathcal{H}(1+\tilde{\omega}_{dm})\frac{\theta_{dm}}{k^2}\right],  \label{eqs1}  \\                                              
\dot{\theta}_{dm}&=-\mathcal{H}(1-3c^2_{s,ad})\theta_{dm}+\frac{k^2c^2_{s,eff}}{1+\tilde{\omega}_{dm}}\delta_{dm}-k^2\sigma_{dm}, \label{eqs2}   \end{align}
where $\delta_{dm}$ and $\theta_{dm}$ are the overdensity and the velocity divergence of DM fluid, $\tilde{\omega}_{dm}\equiv\omega_{dm}(a)$, $\mathcal{H}\equiv aH$ denotes the conformal Hubble parameter, $c^2_{s,eff}$ is the effective sound speed of DM depicting the small-scale behaviors of DM, $c^2_{s,ad}\equiv \tilde{\omega}_{dm}-\dot{\tilde{\omega}}_{dm}/[3\mathcal{H}({1+\tilde{\omega}_{dm}})]$ is the adiabatic sound speed of DM,  and $\sigma_{dm}$ represents the shear perturbation of DM characterizing the anisotropic stress of DM on cosmic scales \cite{Hu:1998kj}. In this study, we use the standard CPL DDE perturbation used by the Planck collaboration \cite{Planck:2018vyg1}.

To investigate quantitatively the LSS behaviors of DDM in theory, we assume the Planck fiducial cosmology, i.e.,  $\Omega_bh^2=0.02237$, $\Omega_ch^2=0.1200$, $100\theta_{\rm MC}=1.04092$, $\tau=0.0544$, $\ln(10^{10}A_s)=3.044$, and $n_s=0.9649$, and take the following two scenarios: (i) varying $\omega_{dm0}$ but fixing $\omega_{dma}=0$; (ii) varying $\omega_{dma}$ but fixing $\omega_{dm0}=0$. In the DDM model, we study the impacts of different parameter values on the LSS quantities of interests including the matter PS, angular PS of CMB, CMB lensing potential PS, Weyl potential PS, velocity PS, density-velocity cross PS, relative baryon-CDM velocity PS, and the cross correlation function between CMB lensing and galaxy number counts.

In Figs.~\ref{fs3} and \ref{fs4}, we show the linear matter PS and angular CMB PS derived from different values of $\omega_{dm0}$, $\omega_{dma}$, $c^2_{s,eff}$, and $\sigma_{dm}$ in the DDM model, respectively. We find that a positive pressure of DM, namely $\omega_{dm0}>0$, lead to an enhancement of structure formation when $k>10^{-3}$ $h$ Mpc$^{-1}$, while negative pressure DM gives a suppression of cosmic structure growth. Interestingly, varying $\omega_{dma}$ provides a similar behavior to the case of varying $\omega_{dm0}$ but the same value of $\omega_{dma}$ as $\omega_{dm0}$ exhibits a smaller deviation from the $\Lambda$CDM model. The small-scale matter clustering is very sensitive to the effective sound speed of DM. A small value of $c^2_{s,eff}$ gives a significant suppression of structure formation when $k>10^{-2}$ $h$ Mpc$^{-1}$. It is interesting that the positive anisotropic stress of DM produces more structures in the range of $k>10^{-1}$ $h$ Mpc$^{-1}$. However, a negative $\sigma_{dm}$ leads to a sudden structure suppression on small scales and the corresponding matter PS finally tends to be close to the case of $\sigma_{dm}>0$ with the same absolute value. For $\sigma_{dm}\sim\mathcal{O}(-3)$, the trough occurs roughly around $k \gtrsim 0.1$ $h$ Mpc$^{-1}$. A smaller $\sigma_{dm}$ will give a smaller suppression on a smaller scale. Furthermore, a positive pressure DM with $\omega_{dm0}>0$ or $\omega_{dma}>0$ gives weaker temperature and E-mode fluctuations than $\Lambda$CDM and shift the CMB spectra towards smaller scales, while a negative pressure DM with $\omega_{dm0}<0$ or $\omega_{dma}<0$ displays stronger temperature and E-mode fluctuations and accordingly shift the whole spectra to larger scales.

\begin{figure}[t!]
	\centering
	\includegraphics[scale=0.58]{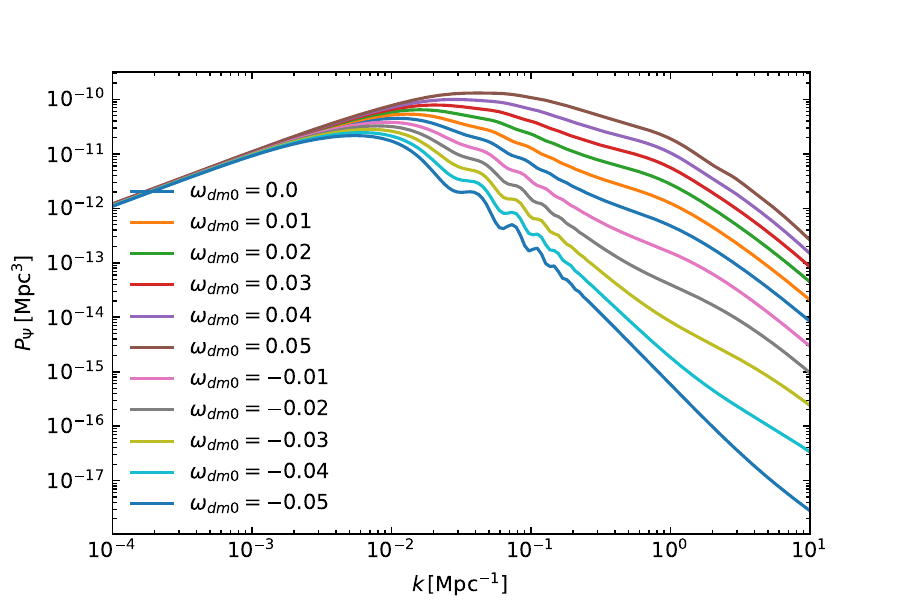}
	\includegraphics[scale=0.58]{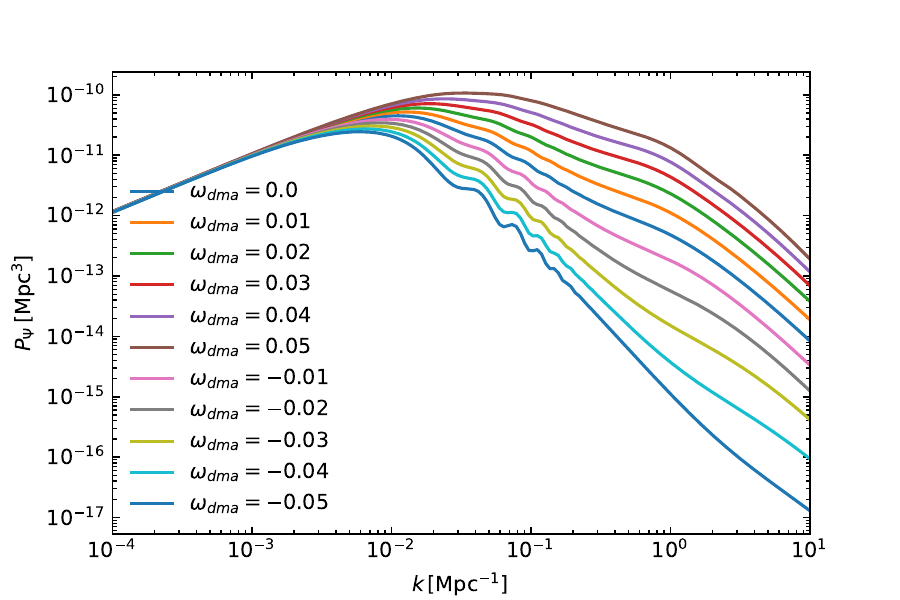}
	\caption{The Weyl potential power spectra derived from different values of $\omega_{dm0}$ and $\omega_{dma}$ in the DDM model, respectively.}\label{fs5}
\end{figure}

\begin{figure}[t!]
	\centering
	\includegraphics[scale=0.58]{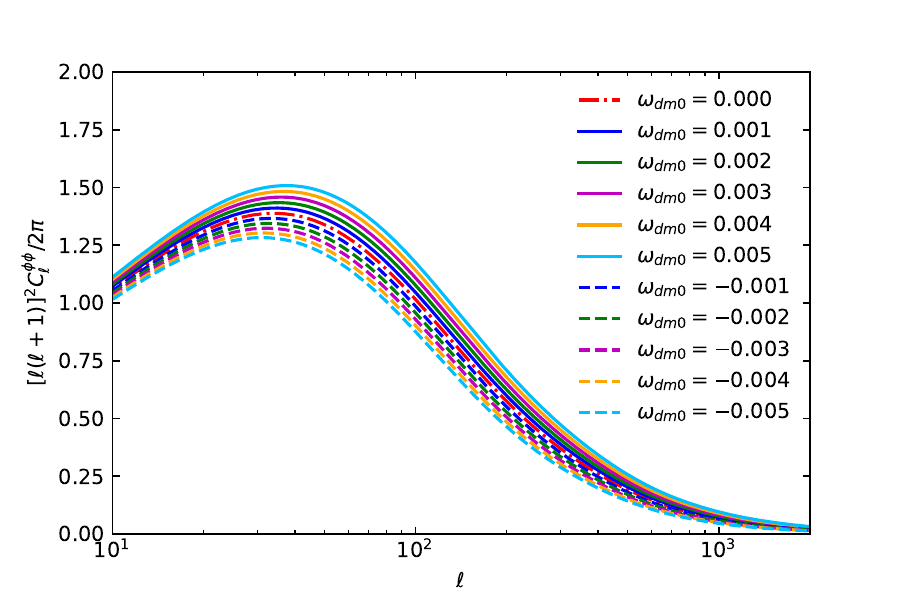}
	\includegraphics[scale=0.58]{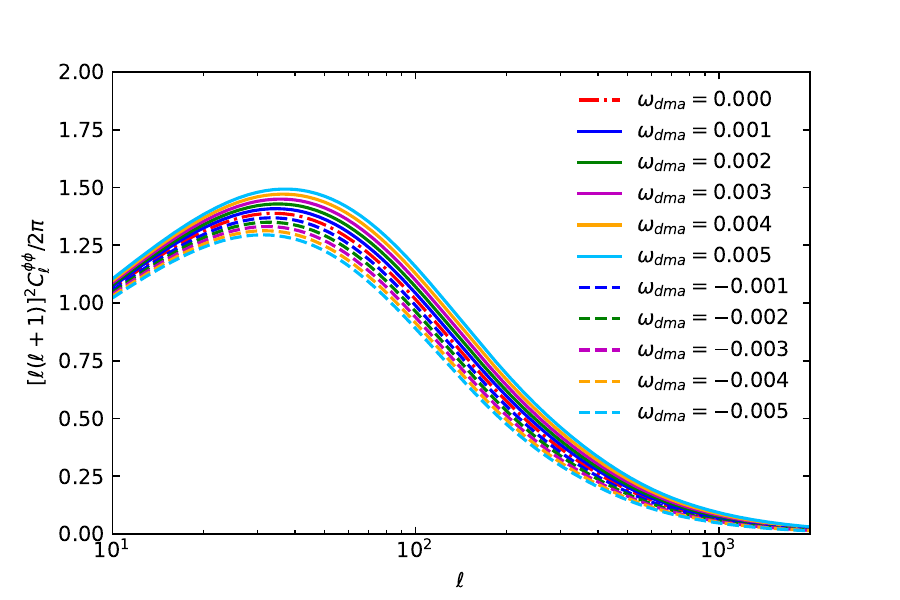}
	\caption{The CMB lensing potential power spectra derived from different values of $\omega_{dm0}$ and $\omega_{dma}$ in the DDM model, respectively.}\label{fs6}
\end{figure}

Generally speaking, in the metric theory of gravity, the Weyl potential $\Psi$ is expressed in terms of two Newtonian-gauge potentials $\phi_N$ and $\psi_N$ by $\Psi = (\psi_N + \phi_N)/2$: it is the scalar field which determines the so-called Weyl tensor (non-local part of the Riemann tensor \cite{Riemann:1854}) in the presence of scalar perturbations \cite{Hanson:2009kr}. In the Fourier space, assuming the linear evolution from the primordial density fluctuation $\mathcal{R}({\bf k})$ with its three-dimensional power spectrum $\mathcal{P}_{\mathcal{R}}({\bf k})$, $\Psi({\bf k})=T_\Psi(k,\eta_0-\chi)\mathcal{R}({\bf k})$, where $T_\Psi(k,\eta_0-\chi)$ denotes the transfer function with $\chi$ and $\eta_0$ being the conformal distance and today's conformal time. The lensing potential reads as
\begin{equation}
\phi({\bf x})=-2\int_{0}^{\chi_\star}\mathrm{d}\chi\left(\frac{\chi_\star-\chi}{\chi_\star\chi}\right)\Psi(\chi{\bf x}, \eta_0-\chi)
\end{equation}
where $\chi_\star$ is the conformal distnace to the last scattering surface. Then, expanding $\phi({\bf x})$ in the harmonic space, the lensing potential PS is easily shown as
\begin{equation}
C_\ell^{\phi\phi}=16\pi\int\frac{\mathrm{d}k}{k}\mathcal{P}_{\mathcal{R}}(k)\left[\int_{0}^{\chi_\star}\mathrm{d}\chi T_\Psi(k,\eta_0-\chi)j_\ell(k\chi)\left(\frac{\chi_\star-\chi}{\chi_\star\chi}\right) \right]^2
\end{equation}
where $j_\ell(k\chi)$ is the spherical Bessel function.

The Weyl potential PS and lensing potential PS derived from different values of $\omega_{dm0}$ and $\omega_{dma}$ in the DDM model are presented in Figs.~\ref{fs5} and \ref{fs6}, respectively. One can easily find that a positive (negative) $\omega_{dm0}$ increases (decreases) the Weyl potential and consequently the lensing potential starting roughly from $k\sim10^{-3}$ $h$ Mpc$^{-1}$ to small scales. Same as above, varying $\omega_{dma}$ gives a similar behavior to the case of varying $\omega_{dm0}$ but the same $\omega_{dma}$ and $\omega_{dm0}$ show a little smaller deviation from the $\Lambda$CDM model.

\begin{figure}[t!]
	\centering
	\includegraphics[scale=0.58]{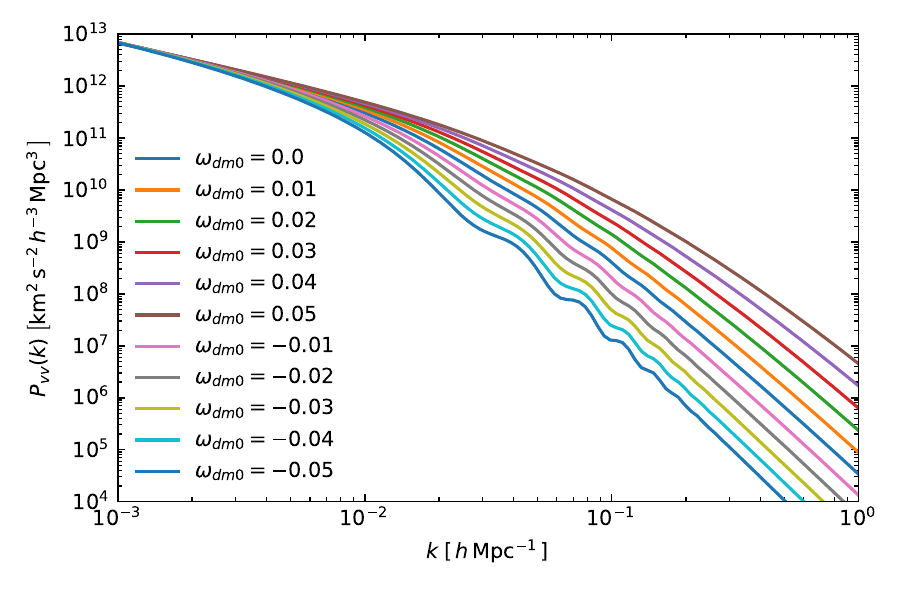}
	\includegraphics[scale=0.58]{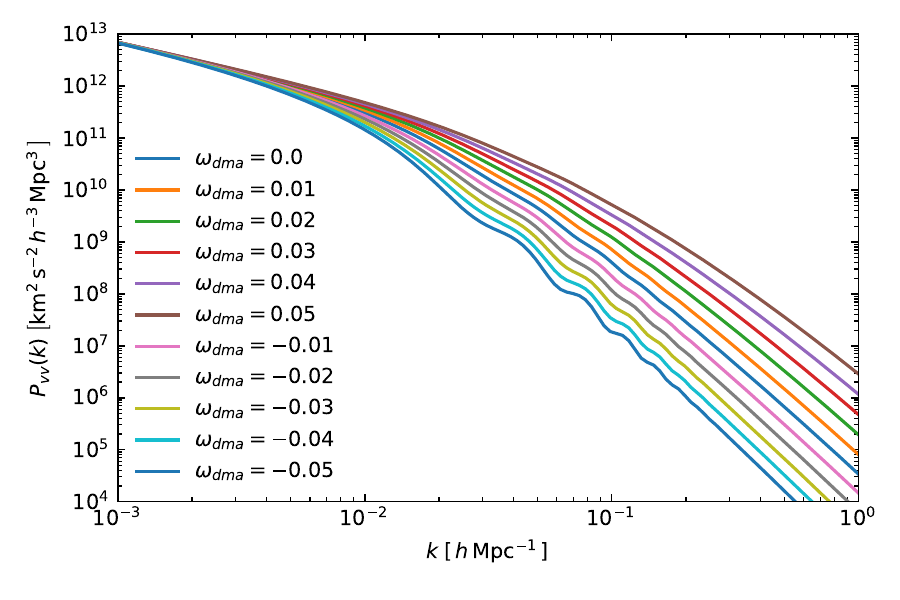}
	\caption{The velocity power spectra derived from different values of $\omega_{dm0}$ and $\omega_{dma}$ in the DDM model, respectively.}\label{fs7}
\end{figure}

\begin{figure}[t!]
	\centering
	\includegraphics[scale=0.52]{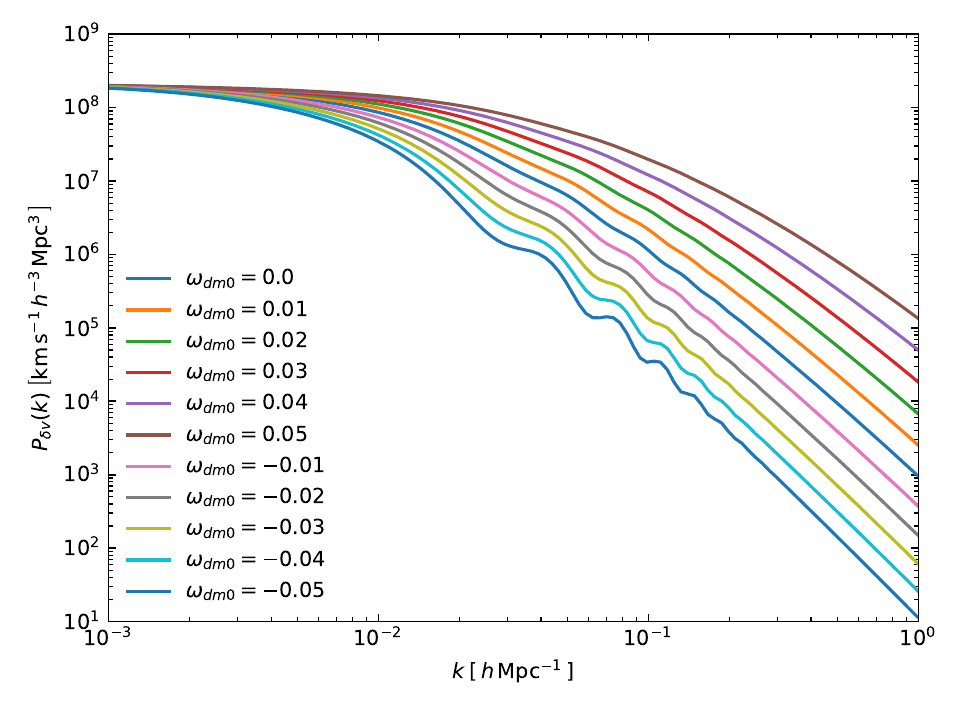}
	\includegraphics[scale=0.52]{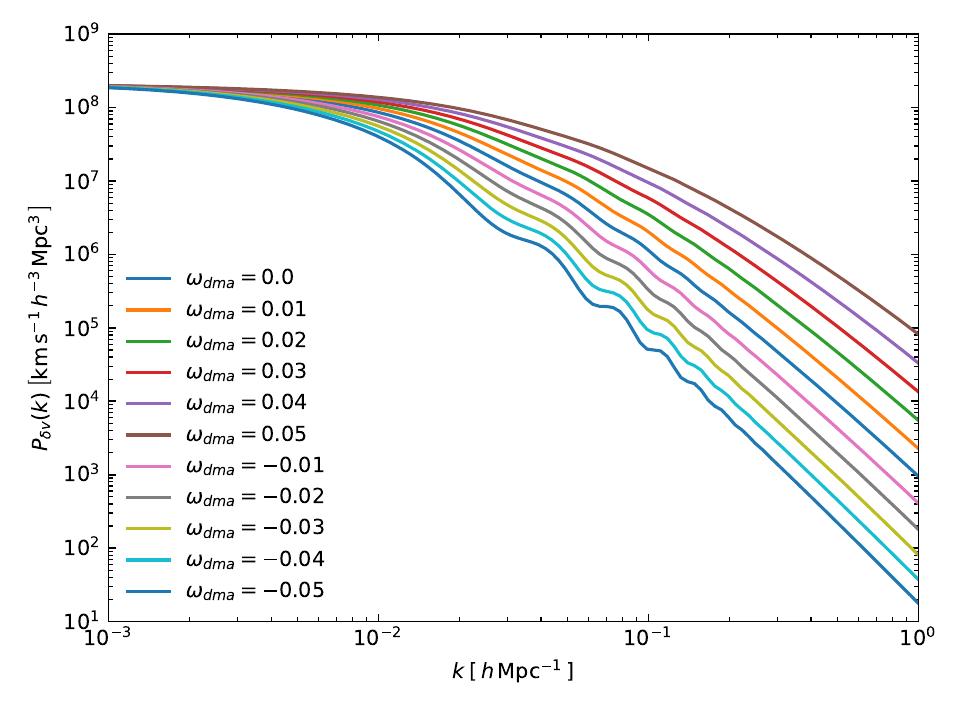}
	\caption{The density-velocity cross power spectra derived from different values of $\omega_{dm0}$ and $\omega_{dma}$ in the DDM model, respectively.}\label{fs8}
\end{figure}

In the linear regime, the peculiar velocity field is related to density perturbations by the continuity equation.
Hence, it is easy to express the velocity PS $P_{vv}(k)$ via the matter PS $P(k)$ as
\begin{equation}
P_{vv}(k)=\left(\frac{f\dot{a}}{k}\right)^2P(k),
\end{equation}
where the linear growth rate $f\equiv\mathrm{d}\ln D/\mathrm{d}\ln a$ and $D$ is the growth factor. Furthermore, the density-velocity cross PS reads as
\begin{equation}
P_{\delta v}(k)=\left(\frac{f\dot{a}}{k}\right)P(k).
\end{equation}
In cosmology, the velocity PS plays a crucial role in understanding the large-scale motion of matter and how structures like galaxies, clusters, and voids form and evolve in the expanding universe. It captures how peculiar velocities --- deviations from the uniform Hubble expansion --- are distributed across different spatial scales.
In Figs.~\ref{fs7} and \ref{fs8}, we present the velocity PS and density-velocity PS derived from different values of $\omega_{dm0}$ and $\omega_{dma}$ in the DDM model, respectively. One can easily find that the DM with positive pressure, i.e., $\omega_{dm0}>0$, gives an enhanced velocity PS on all scales, which could be explained by a possible deeper potential well, a stronger pairwise-kSZ signal, a larger structure growth rate, or the fifth force. For $\omega_{dm0}<0$, we obtain the opposite results. Moreover, we also observe similar behaviors of $\omega_{dm0}$ in the density-velocity cross spectrum. Considering an evolving DM, varying $\omega_{dma}$ only gives a smaller deviation from the $\Lambda$CDM's prediction when $\omega_{dma}=\omega_{dm0}$.

\begin{figure}[t]
	\centering
	\hspace*{-1cm}
	\includegraphics[scale=0.65]{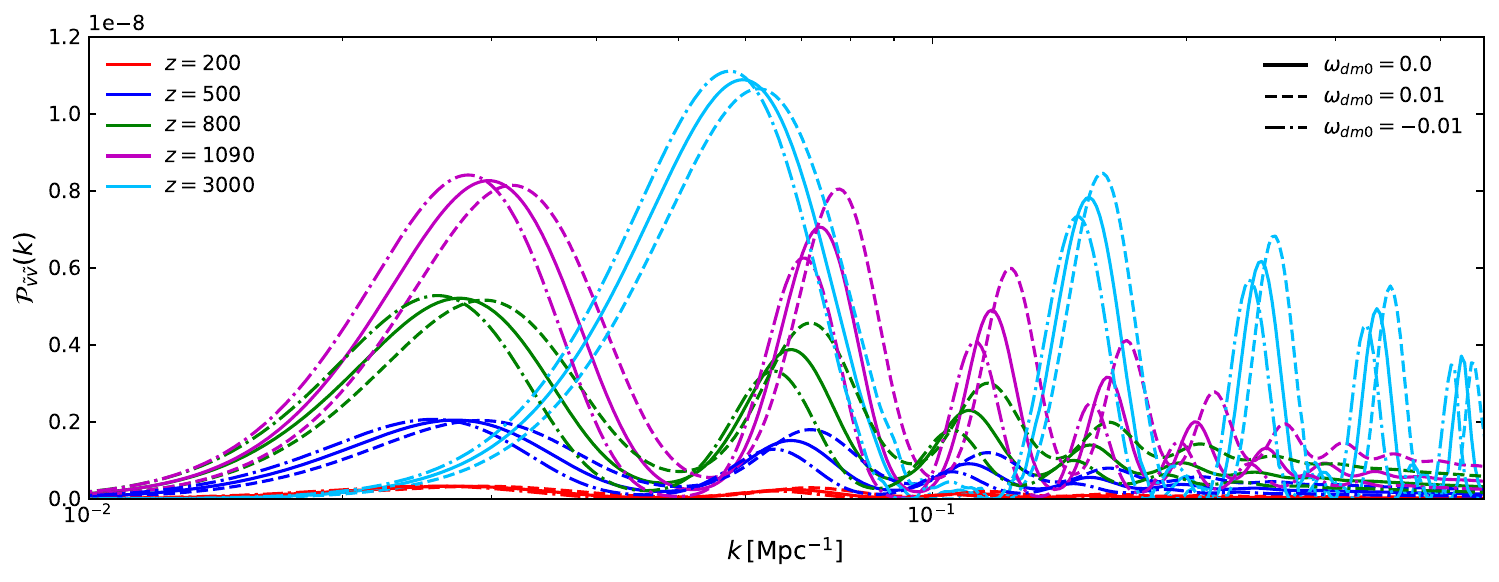}
	\hspace*{-1cm}
	\includegraphics[scale=0.65]{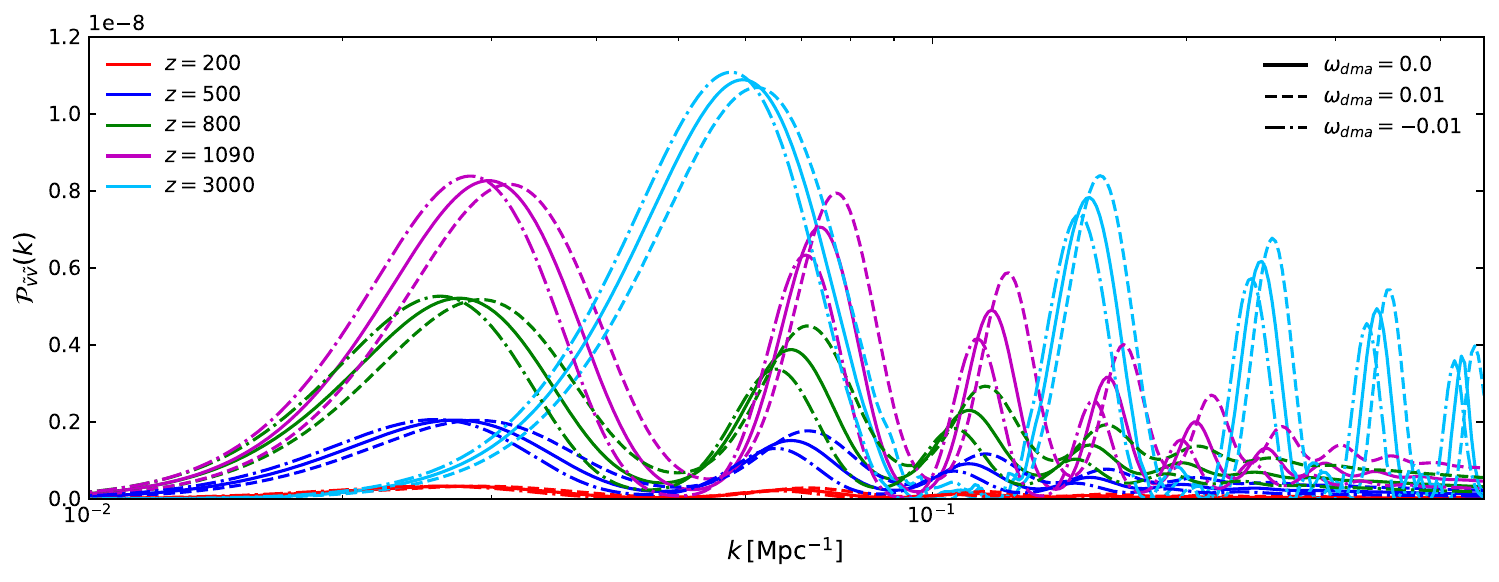}
	\caption{The power spectra of the relative baryon-DM velocity derived from different values of $\omega_{dm0}$ and $\omega_{dma}$ at the redshifts $z=200$, $500$, $800$, $1090$, and $3000$ in the DDM model, respectively.}\label{fs9}
\end{figure}

The relative baryon-DM velocity PS quantifies the statistical properties of the relative velocity field between baryons and DM in the early universe \cite{Tseliakhovich:2010bj}. This concept is important in cosmology, especially in understanding the formation of the first structures such as the Population III stars and the early galaxies. After recombination, baryons decoupled from photons and began to fall into DM potential wells. Nonetheless, because of their prior coupling to radiation, baryons had a residual velocity relative to DM, which can be coherent over large scales (e.g. tens of Mpc). This relative velocity can suppress small-scale structure formation, especially for early star-forming halos. 

The relative velocity field read as follows
\begin{equation}
\tilde{v}({\bf x})=v_b({\bf x})-v_c({\bf x}),
\end{equation}
where $v_b$ and $v_c$ are the velocity fields of baryons and DM, respectively. The corresponding power spectrum $\tilde{P}_{\tilde{v}\tilde{v}}(k)$ is defined as
\begin{equation}
<\tilde{v}_i({\bf k})\tilde{v}^*_j({\bf k'})>=(2\pi^3)\delta^3({\bf k}-{\bf k'})\left(\delta_{ij}-\frac{k_ik_j}{k^2}\right)\tilde{P}_{\tilde{v}\tilde{v}}(k),
\end{equation}
where $i,j=1,2,3$ denote the components of the velocity vector and wavevector in the Cartesian coordinate. The term $\delta_{ij}-\frac{k_ik_j}{k^2}$ projects onto the transverse modes, ensuring the velocity field is divergence-free for the irrotational flow. Furthermore, we define the dimensionless relative baryon-DM velocity PS as 
\begin{equation}
P_{\tilde{v}\tilde{v}}(k)=\frac{k^3}{2\pi^2}\tilde{P}_{\tilde{v}\tilde{v}}(k).
\end{equation}

In Fig.~\ref{fs9}, we present the relative baryon-DM velocity PS derived from different values of $\omega_{dm0}$ and $\omega_{dma}$ in the DDM model, respectively, where the features of BAO are clearly observed. A positive (negative) $\omega_{dm0}$ (or $\omega_{dma}$) can shift the spectrum to smaller (larger) scales, reduce (enhance) the amplitude of the first peak, and enhance (reduce) the amplitudes of the left peaks. In general, this could also be explained by modifying the duration of recombination or cosmic expansion rate in order to alter the acoustic oscillations and the relative velocity field. A faster cosmic expansion could shrink the sound horizon and consequently shift the peaks to smaller scales, while modified ionization history could change the photon-baryon decoupling dynamics, affecting the amplitude and phase of oscillations. Some other alternatives could be non-standard DM physics, modified sound speed, exotic baryon interactions, modified baryon loading, or non-standard neutrino physics. Same as above, a non-zero amplitude of the evolution of DM EoS leads to the similar shape and amplitude of the relative baryon-DM velocity PS but with a slightly smaller deviation from the $\Lambda$CDM's prediction. The degeneracy between different physical mechanisms should be addressed in a future study.

\begin{figure}[t!]
	\centering
	\includegraphics[scale=0.51]{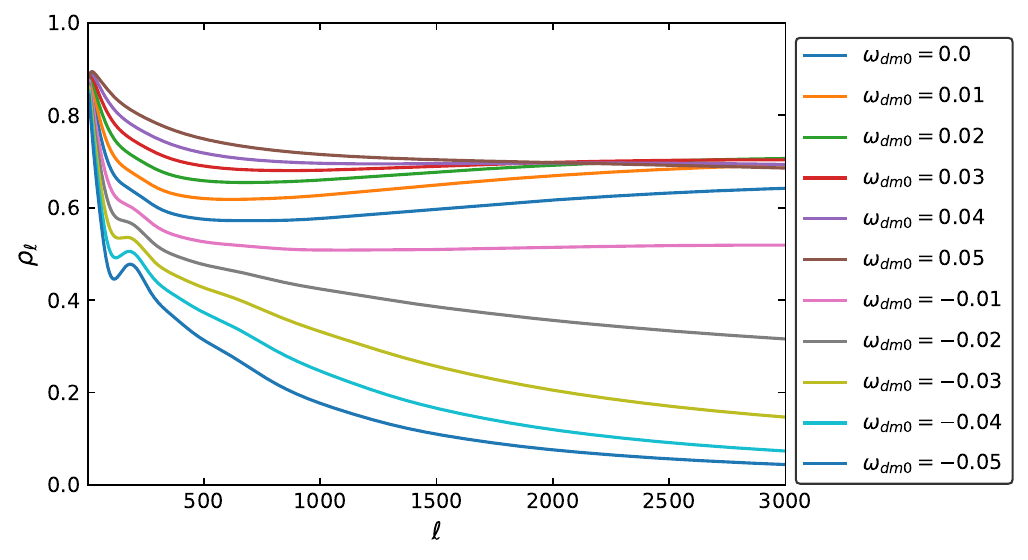}
	\includegraphics[scale=0.51]{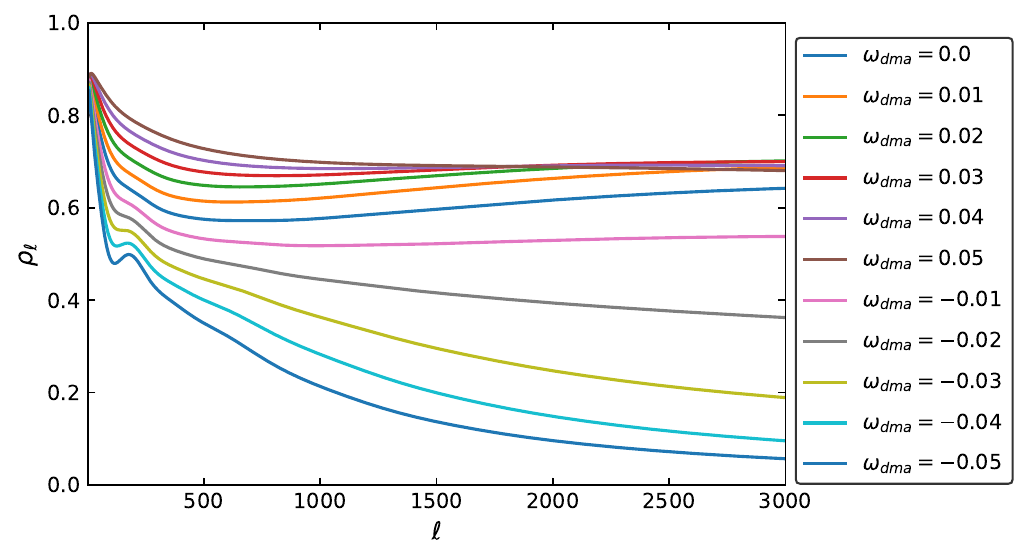}
	\caption{The angular correlation coefficient between the CMB lensing and the simulated galaxy number counts observations of the Vera Rubin Observatory derived from different values of $\omega_{dm0}$ and $\omega_{dma}$ in the DDM model, respectively.}\label{fs10}
\end{figure}

The CMB lensing-galaxy number counts cross-correlation is a powerful cosmological observable that quantifies how the LSS of the universe traced by galaxies correlates with the gravitational lensing of the CMB. Its angular power spectrum is shown as 
\begin{equation}
C_\ell^{\kappa g} \;=\;
\int_{0}^{z_\star} \! \mathrm{d}z\,
\frac{H(z)}{\tilde{\chi}^{2}(z)}
\,W^{\kappa}(z)\,W^{g}(z)\,
P\,\!\bigl(\tfrac{\ell+0.5}{\tilde{\chi}(z)},\,z\bigr).
\end{equation}
where $z_\star$ is the redshift of the last scattering surface, $\tilde{\chi}$ denotes the comoving distance at redshift $z$, and the lensing kernel is 
\begin{equation}
W^\kappa(z) = \frac{3 H_0^2 \, \Omega_m}{2H(z)} \, \frac{\tilde{\chi}(z)}{a(z)} \left[1 - \frac{\tilde{\chi}(z)}{\tilde{\chi}_\star}\right]
\end{equation}
where $\tilde{\chi}_\star$ is the comoving distance to the last scattering surface. The window function of a galaxy survey is 
\begin{equation}
W^g(z)=\frac{b(z)\mathrm{d}N/\mathrm{d}z}{\int(\mathrm{d}N/\mathrm{d}\tilde{z})\mathrm{d}\tilde{z}}
\end{equation}
Similar to the CMB lensing-galaxy number counts cross spectrum, one can easily derive the auto-spectra of two tracers, i.e., $C_\ell^{\kappa\kappa}$ and $C_\ell^{gg}$. Therefore, the angular correlation coefficient $\rho_l$ of the combined
galaxy number counts with CMB lensing observations is easily written as 
\begin{equation}
\rho_l=\frac{C_\ell^{\kappa g}}{\sqrt{C_\ell^{\kappa\kappa}C_\ell^{gg}}}.
\end{equation}
Here we consider an gold sample with i-band magnitude limit $i<25.3$ expected from the Vera Rubin Observatory \cite{RO}. The redshift distribution of these galaxies is approximated by $\mathrm{d}N/\mathrm{d}z=1/(2z_0)(z/z_0)^2 e^{-z/z_0}$, where $z_0 = 0.311$ and $n = 40$ galaxies per squared arcminute. We use a linear redshift dependent galaxy bias $b(z)=1+0.84 z$ for this sample \cite{LSSTScience:2009jmu}. In Fig.~\ref{fs10}, we present the correlation coefficients derived from different values of $\omega_{dm0}$ and $\omega_{dma}$ in the DDM model, respectively. One can find the BAO wiggles imprinted in the cross spectrum, even though they are weaker than in galaxy-galaxy or CMB-only BAO. Interestingly, a positive (negative) $\omega_{dm0}$ (or $\omega_{dma}$) can lead to a stronger (weaker) correlation between the CMB lensing and the galaxy number counts. It is worth noting that a non-zero amplitude of the evolution of DM EoS gives the similar shape and amplitude of the CMB lensing-galaxy number counts cross spectrum but with a slightly smaller deviation from the standard prediction.

\end{document}